\documentclass{article}
\usepackage{amsfonts,amssymb, amsmath}
\usepackage[english]{babel}
\usepackage{graphicx}
\usepackage{float}

\newtheorem{prop}{Proposition}

\newtheorem{ex}{Example}

\def\bq{ \begin{equation}}
\def\eq{ \end{equation}}
\def\ben{ \begin{eqnarray}}
\def\en{ \end{eqnarray}}
\def\a{{\alpha}}

\begin{document}


\title{The Kepler problem: polynomial algebra of non-polynomial first integrals. }

\author{A.V. Tsiganov \\
\it\small St. Petersburg State University, St. Petersburg, Russia\\
\it\small e--mail: andrey.tsiganov@gmail.com}
\date{}
\maketitle

\begin{abstract}
The sum of elliptic integrals simultaneously determines orbits in thr Kepler problem and the addition of divisors on  elliptic curves.  Periodic motion of a body  in physical space is defined by symmetries,  whereas periodic motion of divisors is defined by a fixed point on  the curve. Algebra of the first integrals associated with symmetries is a well-known mathematical object, whereas algebra of the first integrals associated with coordinates of fixed points is unknown. In this paper, we discuss polynomial algebras of non-polynomial first integrals  of superintegrable systems associated with elliptic curves.

 \end{abstract}

\section{Introduction}
\setcounter{equation}{0}
The main point of  interest in integrable systems relies on the fact that they can be integrated by quadratures. For many known integrable systems these quadratures involve various sums of Abelian integrals, which are  inextricably entwined with the arithmetic of divisors. In physics, we describe first integrals of dynamical systems in terms of physical variables, and usually these first integrals are related to symmetries, including dynamical ones. For the Kepler problem the corresponding first integrals are well-known polynomials in momenta \cite{cor03,eul,gul,gg,ks,lag}.

In algebraic geometry we describe the evolution of divisors in terms of coordinates of divisors. The corresponding constants of motion are nothing more than coordinates of fixed points, which are algebraic functions on original physical variables. In fact, algebraic first integrals for the Kepler problem have been obtained by Euler as a by-product of his study of the algebraic orbits appearing in  two fixed centers problem \cite{eul}.

Algebras of the polynomial first integrals of superintegrable systems can be associated with  orthogonal polynomials,  see e.g. \cite{dask1,dask2,vin14,gran1,gran2,kal1,kal2} and references within. For instance, it could be the Racah-Wilson algebra,  Bannai-Ito algebra, Askey-Wilson  algebra, etc. We suppose that the polynomial  algebra of non-polynomial first integrals arising in  divisor arithmetic on elliptic and hyperelliptic curves  may be  associated with elliptic and hyperelliptic functions. It could be Weierstrass functions, Jacobi  functions,  Abelian functions, etc.

In 1762 Euler wrote a paper titled  "\textit{Task: the body is attracted to two given fixed centers inversely proportional to the square of the distance; find in which case the curve described by the body will be algebraic}" \cite{eul}. In this paper he separated algebraic orbits from  transcendental ones using elliptic coordinates  on a plane and an addition law for the corresponding  elliptic integrals.  Algebraic orbits were interesting  because if one of the centers was absent, the body would move on algebraic orbits, as a solution of the Kepler problem.

Indeed, let us consider motion of the body attracted to two fixed centers by forces inversely proportional to the squares of the distance
\[
R=\dfrac{\alpha}{r^2}\qquad\mbox{and}\qquad Q=\dfrac{\beta}{q^2}\,.
\]
In elliptic coordinates $s$ and $u$ equations of motion are reduced to differential equations
\[
\begin{array}{c}
 \dfrac{ds}{\sqrt{Hs^4+(\alpha+\beta)s^3+cs^2-h^2(\alpha+\beta)s-Hh^4-(b^2+c)h^2}}=\\
 = \dfrac{du}{\sqrt{Hu^4+(\alpha-\beta)u^3+cs^2-h^2(\alpha-\beta)u-Hh^4-(b^2+c)h^2}}
 \end{array}
\]
and
\[
\begin{array}{c}
dt= \dfrac{s^2ds}{4\sqrt{Hs^4+(\alpha+\beta)s^3+cs^2-h^2(\alpha+\beta)s-Hh^4-(b^2+c)h^2}}-\\
 - \dfrac{u^2du}{4\sqrt{Hu^4+(\alpha-\beta)u^3+cs^2-h^2(\alpha-\beta)u-Hh^4-(b^2+c)h^2}}
 \end{array}
\]
which we copied from  page 106 of  Lagrange's textbook \cite{lag}. Here $H$ and $h$ are integrals of motion, which are second order polynomials in momenta,  $b$ and $c$ are geometric parameters describing positions of fixed centers.

 At $\beta=0$ and $x_1=s$, $x_2=u$  these equations become well-known Abel's quadratures for the two-body Kepler problem
\bq\label{ab-eq2}
\int \dfrac{dx_1}{\sqrt{f(x_1)}}+ \int \dfrac{dx_2}{\sqrt{f(x_2)}}=const\,,
\eq
and
\bq\label{ab-eq1}
 \int \dfrac{x_1^2dx_1}{\sqrt{f(x_1)}}+ \int \dfrac{x_2^2dx_2}{\sqrt{f(x_2)}}=4t\,,
\eq
on elliptic curve $X$ defined by an equation of the form
\bq\label{ell-eq}
X:\qquad \Phi(x,y)= y^2-f(x)=0\,,\qquad f(x)=a_4x^4+a_3x^3+a_2x^2+a_1x+a_0
\eq
 on a projective plane.  The first equation (\ref{ab-eq2}) determines trajectories of motion, whereas the second  equation (\ref{ab-eq1}) defines time \cite{lag}.

Equations (\ref{ab-eq2}-\ref{ab-eq1}) describe motion of a body in the Kepler problem and, simultaneously, evolution of  points $P_1(t)=(x_1,y_1)$, $P_2(t)=(x_2,y_2)$  around  fixed point $P_3=(x_3,y_3)$ on $X$ governed by arithmetic equation
\bq\label{div-eq}
P_1(t)+P_2(t)=P_3\,.
\eq
 According  to Abel's theorem \cite{ab,bliss24,hens, grif04,gr} trajectories of points $P_{1,2}(t)$ on $X$ are uniquely determined by  Abel's sum (\ref{ab-eq2}) in the same way as trajectories of a body in the Kepler problem. Subsequently, periodic  motion  of  points  along  plane curve $X$ generates periodic  motion in phase space of the Kepler  system and vice versa.

According to \cite{eul0} coordinates of the fixed point $x_3$ and $y_3$ are algebraic functions on the coordinates of movable points $x_{1,2}$ and $y_{1,2}$ which are constants of divisor motion along elliptic curve $X$ (\ref{div-eq}). These algebraic functions on elliptic coordinates $u_{1,2}$ and momenta $p_{u_{1,2}}$ are also first integrals in the Kepler problem. In \cite{eul} Euler used these algebraic first integrals and their combination
\bq\label{eul-int}
C=2a_4x_3^2+a_3x_3+a_2-2\sqrt{a_4}y_3
\eq
in order to separate algebraic orbits from  transcendental ones in the problem of two fixed centers.

In the Kepler case first integral $C$ (\ref{eul-int}) is  a square of the component of angular momentum
\[C=-\bigl(p_1q_2-(q_1-\kappa)p_2\bigr)^2\,,\]
Of course, this first integral is related to the well-studied rotational symmetry \cite{cor03,gul,gg,ks}. The Poisson algebras of polynomial first  integrals for the Kepler problem and other dynamical systems, separable in elliptic, parabolic and polar coordinates are well studied objects, see e.g. \cite{dask1,dask2,vin14,gran1,gran2,kal1,kal2}.

 Our aim is to calculate the algebra of non-polynomial first integrals $x_3$ and $y_3$  and to discuss various representations of this algebra. This algebra occurs in a standard arithmetic of divisors on elliptic curves and, therefore, it could belong to a family of algebras associated with the arithmetic of divisors on more complicated hyperelliptic curves.

 \section{The Kepler problem}
 \setcounter{equation}{0}
In the original physical problem configuration space is 6-dimensional, and phase space is 12-dimensional space.
Discussion of the traditional topics, such as symmetries, conservation of angular momentum, conservation of Laplace-Runge-Lenz vector, regularization and so on, may be found in \cite{cor03,gul,gg,ks} and many other papers and textbooks.

Our aim is to come back to Euler's calculations in order to get  a family of superintegrable systems with nonpolynomial first integrals, which cannot be obtained using symmetries.  Following  Euler \cite{eul} we start with the planar two centers  problem. Reduction of the original phase space to the orbital plane, which Euler described by using a picture,  may be found in the Lagrange textbook \cite{lag}.

 \subsection{Motion in orbit}
 Let us introduce elliptic coordinates on the orbital plane.  If  $r$ and $r'$ are distances  from a point on the plane to the  fixed centers, then elliptic coordinates  $u_ {1,2} $ are
\[
r+r'=2u_1\,,\qquad r-r'=2u_2\,.
\]
If the centres are taken to be fixed at $-\kappa$ and $\kappa$ on   $OX$-axis of the Cartesian coordinate system, then we have standard Euler's definition of  elliptic coordinates on the plane
\bq
\label{ell-coord}
q_1 =\dfrac{u_1u_2}{\kappa}\,,\qquad \mbox{and}\qquad q_2 = \dfrac{\sqrt{(u_1^2-\kappa^2)(\kappa^2-u_2^2)}}{\kappa}\,.
\eq
Coordinates $u_{1,2}$ are curvilinear orthogonal coordinates, which take values only  in the intervals
\[u_2<\kappa<u_1\,,\]
i.e. they are locally defined coordinates. The corresponding momenta are given by
\bq\label{ell-mom}
\begin{array}{rcl}
p_1&=& \dfrac{u_1 u_2 (p_{u_1} u_1-p_{u_2} u_2)-\kappa^2(p_{u_1} u_2-p_{u_2} u_1)  }{\kappa(u_1^2- u_2^2)}\,,\\ \\
p_2&=& \dfrac{(p_{u_1} u_1-p_{u_2} u_2)\sqrt{u_1^2-\kappa^2}\sqrt{\kappa^2-u_2^2} }{\kappa(u_1^2-u_2^2)}\,.
\end{array}
\eq
For the planar Kepler problem with one center of attraction at  point $(\kappa,0)$, which is  a partial case of Euler's two-centers problem,   Hamiltonian and first integral are equal to
 \bq\label{int-kepl0}
 2H=I_1=p_1^2+p_2^2+\dfrac{\a}{r}\,,\qquad
 I_2= \dfrac{\a (r^2-r'^2)}{4r}-(\kappa^2+q_2^2)p_1^2-2q_1q_2p_1p_2-q_1^2p_2^2\,.
 \eq
 In elliptic coordinates these integrals of motion have the following form
  \bq\label{int-kepl1}
 \begin{array}{rcl}
 I_1&=&\dfrac{(u_1^2-\kappa^2)p_{u_1}^2}{u_1^2-u_2^2}+
 \dfrac{(u_2^2-\kappa^2)p_{u_2}^2}{u_2^2-u_1^2}
  +\dfrac{\a}{u_1+u_2}\\
 \\
I_2&=&\dfrac{u_2^2(u_1^2-\kappa^2)p_{u_1^2}}{u_2^2-u_1^2}
 +\dfrac{u_1^2(u_2^2-\kappa^2)p_{u_2}^2}{u_1^2-u_2^2}+ \dfrac{\alpha u_1u_2}{u_1+u_2}\,.
 \end{array}
 \eq
 Substituting solutions of these equations with respect to $p_{u_1}$ and $p_{u_2}$ into the equations of motion
 \[
 \dfrac{d{u}_1}{dt}=\{u_1,H\}=\dfrac{(u_1^2-\kappa^2)p_{u_1}}{u_1^2-u_2^2}\,,\qquad
  \dfrac{d{u}_2}{dt}=\{u_2,H\}=\dfrac{(u_2^2-\kappa^2)p_{u_2}}{u_2^2-u_1^2}\,,
 \]
 we obtain differential equations of the form
 \[\begin{array}{rcl}
 \dfrac{du_1}{\sqrt{(u_1^2-\kappa^2) (I_1u_1^2-\a u_1+I_2)}}&=&\phantom{-}\dfrac{dt}{u_1^2-u_2^2}\,,\\
 \\
 \dfrac{du_2}{\sqrt{(u_2^2-\kappa^2) (I_1u_2^2-\a u_2+I_2)}}&=&-\dfrac{dt}{u_1^2-u_2^2}\,,
 \end{array}
 \]
 After integration of the sum of these equations one gets  a sum of  Abelian integrals
 \bq\label{ell-sum-11}
  \int \dfrac{du_1}{\sqrt{(u_1^2-\kappa^2) (I_1u_1^2-\a u_1+I_2)}} + \int \dfrac{du_2}{\sqrt{(u_2^2-\kappa^2) (I_1u_2^2-\a u_2+I_2)}}=const
  \eq
 involving holomorphic differentials on elliptic curve $X$  (\ref{ell-eq}) defined by equation
 \bq\label{ell-kepl}
X:\quad\Phi(x,y)= y^2-f(x)=0\,,\qquad f(x)=I_1x^4-\alpha x^3+(I_2-I_1\kappa^2)x^2+\kappa^2\alpha x-I_2\kappa^2\,.
   \eq
Here $I_{1,2}$ are values of  the integrals of motion, see terminology and discussion  in  Lagrange's textbook \cite{lag} and in comments by Darboux and Serret  \cite{darb,ser}.

 \subsection{Motion in elliptic curve}
Using the sum of Abelian integrals (\ref{ell-sum-11})  we can transfer from classical mechanics to algebraic geometry and, in particular, to divisors arithmetic  on elliptic curve. Indeed, coordinates of movable points $P_{1,2}(t)$ in the equation of motion along elliptic curve $X$ (\ref{div-eq}) are
\[x_1=u_1,\quad y_1=(u_1^2-\kappa^2)p_{u_1}\qquad\mbox{and}\qquad x_2=u_2\,,\quad y_2=(u_2^2-\kappa^2)p_{u_2}\,.\]
Because
\[
u_2<\kappa<u_1\quad \Rightarrow \quad x_1\neq x_2\quad \Rightarrow \quad (x_3,y_3)\neq (\infty,\infty)\,,
\]
abscissa  $x_3$ and ordinate $y_3$ of fixed point $P_3$ are well-defined finite functions on $T^*\mathbb R^2$.

In order to calculate affine coordinates of fixed point $P_3$ we have to consider intersection of $X$ and parabola $Y$ with a fixed leading coefficient
\[
Y:\qquad y=\mathcal P(x)\,,\qquad \mathcal P(x)=\sqrt{a_4}x^2+b_1x+b_0\,,
\]
see \cite{ab,grif04,gr} for  details. Solving equations
\[
y_1=\sqrt{a_4}x_1^2+b_1x_1+b_0\qquad\mbox{and}\qquad  y_2=\sqrt{a_4}x_2^2+b_1x_2+b_0
\]
with respect to $b_1$ and $b_0$  we calculate standard interpolation by Lagrange for polynomial 
 \bq\label{int-par}
 \mathcal P(x)=\sqrt{a_4}x^2+b_1x+b_0=\sqrt{a_4}(x_1-x)(x_2-x)+\dfrac{ (x-x_2)y_1}{x_1-x_2}
 +\dfrac{(x-x_1)y_2}{x_2-x_1}\,.
 \eq
Substituting $y=\mathcal P(x)$ into $f(x)-y^2=0$ we obtain Abel's polynomial
\[\begin{array}{rcl}
\psi&=&f(x)-\mathcal P^2(x)\\ \\
&=&(a_3-2b_1\sqrt{a_4})x^3+(a_2-2b_0\sqrt{a_4}-b_1^2)x^2+(a_1-2b_0b_1)x+a_0-b_0^2\\ \\
&=&(a_3-2b_1\sqrt{a_4})(x-x_1)(x-x_2)(x-x_3)\,.
\end{array}
\]
Evaluating coefficients of this polynomial we determine  abscissa of the fixed point $P_3$ in (\ref{div-eq})
\bq\label{coord-x3}
x_3=-x_1-x_2-\dfrac{2b_0\sqrt{a_4}+b_1^2-a_2}{2b_1\sqrt{a_4}-a_3}
\eq
and its ordinate
\bq\label{coord-y3}
y_3=-\mathcal P(x_3)=-\sqrt{a_4}x_3^2-b_1x_3-b_0\,,
\eq
where $b_1$ and $b_0$ are functions of coordinates of movable points  $x_1,x_2$ and $y_1,y_2$ defined by equation (\ref{int-par}).

Now we  come back from divisor arithmetics to classical mechanics. For the Kepler problem we have
\[
a_4=I_1,\quad a_3=-\alpha\,,\quad a_2=(I_2-\kappa^2I_1)\,,\quad
a_1=\kappa^2\alpha\,,\quad a_0=-\kappa^2 I_2\,,
\]
so abscissa of fixed point $P_3$ is equal to
\[
x_3=\scriptstyle \frac{2 \bigl(\kappa^2(p_{u_1} u_2-p_{u_2} u_1)-u_1u_2(p_{u_1} u_1+p_{u_2} u_2)\bigr)\bigl(\sqrt{I_1}(u_1^2-u_2^2)-\kappa^2(p_{u_1}- p_{u_2})+p_{u_1} u_1^2-p_{u_2} u_2^2\bigr)
- \alpha(u_1-u_2)^2 (\kappa^2+u_1 u_2)}{\left(u_1^2-u_2^2\right)
\bigl(
2\sqrt{I_1}\left(\kappa^2(p_{u_1}-p_{u_2})-p_{u_1} u_1^2+p_{u_2} u_2^2\right)+
2\kappa^2(p_{u_1}^2-p_{u_2}^2)-2p_{u_1}^2u_1^2+2p_{u_2}^2u_2^2-
\alpha (u_1-u_2)\bigr)}\,.
\]
Ordinate $y_3$ (\ref{int-par}) is equal to
\[y_3=-\sqrt{I_1}(u_1-x_3)(u_2-x_3)-\dfrac{ (x_3-u_2)(\kappa^2-u_1^2)p_{u_1}}{u_1-u_2}
 -\dfrac{(x_3-u_1)(\kappa^2-u_2^2)p_{u_2}}{u_2-u_1}\,,
\]
Here $I_1$ is given by (\ref{int-kepl1}) and, therefore, $x_3$ and $y_3$ are algebraic functions on $u_{1,2}$ and $p_{u_{1,2}}$.

In \cite{eul,eul0} Euler introduced  algebraic first integral  $C$ (\ref{eul-int}) which   is nothing more than a square of angular momentum  in the Kepler case:
\[\begin{array}{rcl}
C&=&2a_4x_3^2+a_3x_3+a_2-2\sqrt{a_4}y_3=\dfrac{(u_1^2-\kappa^2)(u_2^2-\kappa^2)(p_{u_1}-p_{u_2})^2}{(u_1-u_2)^2}=\\ \\
&=&-\left(p_1q_2-(q_1-\kappa)p_2\right)^2\,.
\end{array}
\]
It is well-known that existence of this first integral $C$ is related to rotational symmetry of the orbital plane a around center of attraction.
Algebraic first integrals $x_3$ and $y_3$ have no obvious physical meaning, but they have a trivial geometric description as  affine coordinates of the fixed point on elliptic curve $X$.

\subsection{Symmetry breaking}
Let us consider  non-canonical transformations of momenta preserving symmetries of configuration space, but breaking symmetry between divisors  \cite{ts18,ts08,ts08a, ts09, ts10, ts18s, ts19s}.

It is easy to see, that transformation of momenta
\bq\label{trans-mom}
p_{u_1}\to \dfrac{p_{u_1}}{m}\qquad\mbox{and}\qquad p_{u_2}\to \dfrac{p_{u_2}}{n}\,,
\eq
where $m$ and $n$ are rational numbers,  preserves the only symmetry of potential part of first integrals  and breaks symmetry of whole integrals of motion (\ref{int-kepl0}), which  now have the form
 \bq\label{int-kepl-mn}
 \begin{array}{rcl}
2H= I_1&=&\dfrac{u_1^2-\kappa^2}{u_1^2-u_2^2}\left(\dfrac {p_{u_1}}{m}\right)^2+
 \dfrac{u_2^2-\kappa^2}{u_2^2-u_1^2}\left(\dfrac {p_{u_2}}{n}\right)^2
  +\dfrac{\a}{u_1+u_2}\\
 \\
I_2&=&\dfrac{u_2^2(u_1^2-\kappa^2)}{u_2^2-u_1^2}\left(\dfrac {p_{u_1}}{m}\right)^2
 +\dfrac{u_1^2(u_2^2-\kappa^2)}{u_1^2-u_2^2}\left(\dfrac {p_{u_2}}{n}\right)^2+ \dfrac{\alpha u_1u_2}{u_1+u_2}\,.
 \end{array}
 \eq
 In Cartesian coordinates on the  plane  Hamiltonians (\ref{int-kepl-mn})  read as
\[
H=\dfrac{(m^2+n^2)(p_1^2+p_2^2)}{4 m^2 n^2}
+\dfrac{(m^2-n^2)\Bigl( (\kappa^2-q_1^2+q_2^2) (p_1^2-p_2^2)+4 q_1 q_2 p_1 p_2\Bigr)}{4 m^2 n^2 r r'}+\dfrac{\alpha}{2r}\,.
\]
 According to \cite{ts18s, ts19s} these Hamiltonians (\ref{int-kepl-mn})  are superintegrable Hamiltonians because this non-canonical transformation  sends  the original sum of elliptic integrals (\ref{ell-sum-11}) to the sum
 \bq\label{ell-sum-mn}
    m \int \dfrac{du_1}{\sqrt{(u_1^2-\kappa^2) (I_1u_1^2-\a u_1+I_2)}} + n\int \dfrac{du_2}{\sqrt{(u_2^2-\kappa^2) (I_1u_2^2-\a u_2+I_2)}}=const\,,\eq
 i.e. to the sum of elliptic integrals with integer coefficients
\[m_1n_2 \int \dfrac{du_1}{\sqrt{(u_1^2-\kappa^2) (I_1u_1^2-\a u_1+I_2)}} + n_1m_2\int \dfrac{du_2}{\sqrt{(u_2^2-\kappa^2) (I_1u_2^2-\a u_2+I_2)}}=const\,.\]
Here we present rational numbers  $m=m_1/m_2$ and $n=n_1/n_2$ as the ratio of integer numbers.  The corresponding first integrals of motion on elliptic curve $X$ were obtained in Problem 83 of  Euler's textbook  \cite{eul0}.

Without loss of generality  below we  consider only positive integer numbers $m$ and $n$. In this case sum of elliptic integrals  (\ref{ell-sum-mn}), which generates the well-studied arithmetic equation for divisors on elliptic curves
\bq\label{div-eq-mn}
[m]P_1(t)+[n]P_2(t)=P_3\,,
\eq
 see \cite{eul0,lang78, sil}. Here $[k]P$ means scalar multiplication of point on an elliptic curve  on integer number $k\in \mathbb Z$, and we denote coordinates of $[k]P=[k](x, y)$ as $([k]x,[k]y)$, whereas notations for coordinates of $P_3$ in (\ref{div-eq-mn}) remain the same  $x_3$  and $y_3$.

In order to get coordinates of fixed point $P_3$ in  (\ref{div-eq-mn})  we have to:
\begin{enumerate}
  \item Multiply divisors $P_{1,2}$  by integer numbers $m$ and $n$ using a recursion procedure proposed by Euler \cite{eul0}  or using standard expressions for scalar multiplication on elliptic curves, see \cite{lang78, sil,ts18d} and references within.
  \item Add divisors $[m]P_1$ and $[n]P_2$. Because points $ [m] P_1$ and $ [n] P_2$ belong to intersection divisor of $X$ and $Y$, we can use the equation of parabola $Y$
  \[
[m]y_1=\sqrt{a_4}\cdot\left([m]x_1\right)^2+b_1\cdot[m]x_1+b_0\qquad\mbox{and}\qquad [n]y_2=\sqrt{a_4}\cdot\left([n]x_2\right)^2+b_1\cdot[n]x_2+b_0\,.
\]
in order to calculate its coefficients $b_1$ and $b_0$. After that we  substitute $a_i$, $b_i$ and $[m]x_1$, $[m]y_1$ and
$[n]x_2$, $[n]y_2$ into (\ref{coord-x3}) and (\ref{coord-y3}) and obtain coordinates of fixed point $P_3$ in (\ref{div-eq-mn}) :
\bq\label{xy3-mn}
x_3=-[m]x_1-[n]x_2-\dfrac{2b_0\sqrt{a_4}+b_1^2-a_2}{2b_1\sqrt{a_4}-a_3}\,,\qquad
y_3=-\sqrt{a_4}x_3^2-b_1x_3-b_0\,,
\eq
Following \cite{eul0} we can also determine Euler's first integral  of equation of motion (\ref{div-eq-mn}) on $X$:
 \bq\label{eul-int-nm}
 \begin{array}{rcl}
  C_{mn}&=&
 2a_4x_3+a_3x_3+a_2-2\sqrt{a_4}y_3=\\ \\
 &=&
  \left(\dfrac{[m]y_1-[n]y_2}{[m]x_1-[n]x_2} \right)^2-a_4\bigl([m]x_1+[n]x_2\bigr)^2-a_3\bigl([m]x_1+[n]x_2\bigr)\,.
 \end{array}
 \eq
  \item Identify affine coordinates on the projective plane  with elliptic coordinates on phase space
   \bq\label{coord-xy-mn}
 x_1=u_1,\quad y_1=(u_1^2-\kappa^2)\dfrac{p_{u_1}}{m}\qquad\mbox{and}
 \qquad x_2=u_2\,,\quad y_2=(u_2^2-\kappa^2)\dfrac{p_{u_2}}n
 \eq
 so that constants of divisor motion on elliptic curve $X$ become first integrals of Hamiltonian vector field in
 $T^*\mathbb R^2$.
\end{enumerate}

At $m=n$ first integral $C_{mn}$ (\ref{eul-int-nm}) is  a square of angular momentum relating with rotational symmetry. At  $m\neq n$ all first integrals $x_3,y_3$ (\ref{xy3-mn}) and $C_{mn}$ (\ref{eul-int-nm}) are algebraic functions in phase space. Some explicit expressions  of these first integrals may be found in \cite{ts19s} .

Now we are ready to formulate the main result in this note.
\begin{prop}
Functions  $I_1,I_2$ (\ref{int-kepl-mn}) and $x_3,y_3$ (\ref{xy3-mn}) in phase space $T^*\mathbb R^2$
can be considered as representation of the following algebra of the first integrals
\bq\label{alg-int-kepl}
\begin{array}{lll}
 \{I_1,I_2\}=0\,,\qquad &\{I_1,x_3\}=0\,,\qquad &\{I_1,y_3\}=0\,,\\ \\
 \{I_2,x_3\}=\Phi_y(x_3,y_3)\,,\qquad &\{I_2,y_3\}= -\Phi_x(x_3,y_3)\,,\qquad &\{x_3,y_3\}=\kappa^2-x_3^2
 \end{array}
\eq
labelled by two integer numbers $m$ and $n$.  Here
\[
\Phi_y(x,y)=\dfrac{\partial \Phi(x,y)}{\partial y}=2y\quad\mbox{and}\quad
\Phi_x(x,y)=\dfrac{\partial \Phi(x,y)}{\partial x}=-(4I_1x^3-3\alpha x^2+2(I_2-\kappa I_1)x+\alpha\kappa^2)
  \]
are derivatives of  function $\Phi(x,y)$ from the definition of elliptic curve $X$ (\ref{ell-kepl}), and $\{.,.\}$
is the standard canonical Poisson bracket
\[
\{u_1,u_2\}=0\,,\quad \{p_{u_1},p_{u_2}\}=0\,,\quad
\{u_i,p_{u_j}\}=\delta_{ij}\,.
\]
\end{prop}
The  Poisson brackets (\ref{alg-int-kepl})  are derived from the brackets
 \[
\{I_i,I_j\}=\{\omega_i,\omega_j\}=0\,,\qquad \{I_i,\omega_j\}=\delta_{ij}\,,
\]
between  action variables $I_{1,2}$ (\ref{int-kepl-mn})  and angle variables
 \[\begin{array}{rcl}
 \omega_1&=&\displaystyle m \int \dfrac{u_1^2du_1}{\sqrt{(u_1^2-\kappa^2) (I_1u_1^2-\a u_1+I_2)}} + n\int \dfrac{u_2^2du_2}{\sqrt{(u_2^2-\kappa^2) (I_1u_2^2-\a u_2+I_2)}}\,,\\\
    \\
 \omega_2&=&\displaystyle m \int \dfrac{du_1}{\sqrt{(u_1^2-\kappa^2) (I_1u_1^2-\a u_1+I_2)}} + n\int \dfrac{du_2}{\sqrt{(u_2^2-\kappa^2) (I_1u_2^2-\a u_2+I_2)}}\,.
 \end{array}
\]
Here we use indefinite integrals determined only up to an additive constant following  Euler \cite{eul0}, Abel \cite{ab}, Jacobi \cite{jac32} and St\"{a}ckel \cite{st95}, see also discussion in \cite{bliss24,grif04}.

In (\ref{alg-int-kepl}) form of the bracket $\{x_3,y_3\}$  coincides with the form of original brackets $\{x_1,y_1\}$ and $\{x_2,y_2\}$.
Two remaining non-trivial brackets can be rewritten in the following form
\bq \label{eq-mn-ham}
\{\Phi(x,y),x_3\}= \dfrac{\partial \Phi}{\partial I_2} \left.\dfrac{\partial \Phi}{\partial y}\right|_{P_3}\quad\mbox{and}\quad
\{\Phi(x,y),y_3\}=-\dfrac{\partial \Phi}{\partial I_2} \left.\dfrac{\partial \Phi}{\partial x}\right|_{P_3}\,,
\eq
which is reminiscent of Hamiltonian equations of motion. The first time brackets   (\ref{alg-int-kepl})
appeared when we studied superintegrable systems associated with elliptic curve in the short Weierstrass form \cite{ts19s,ts19all}. Below we discuss similar algebras of the first integrals for  other superintegrable systems associated with elliptic curve.

\section{Harmonic oscillator}
Let us consider 2D harmonic oscillator with the following Hamiltonian and additional integral of motion
\[
2H=I_1=p_1^2+p_2^2-\alpha^2(q_1^2+q_2^2)\,,\qquad
I_2=(p_1^2-\alpha^2q_1^2)\kappa^2-(p_1q_2+p_2q_1)^2\,,
\]
which is a shifted square of angular momentum. In elliptic coordinates (\ref{ell-coord}) and (\ref{ell-mom}) these constants of motion are equal to
\bq\label{int-osc1}
\begin{array}{rcl}
I_1&=&\dfrac{(u_1^2-\kappa^2)p_{u_1}^2}{u_1^2-u_2^2}
+\dfrac{(u_2^2-\kappa^2)p_{u_2}^2}{u_2^2-u_1^2}+\alpha^2(\kappa^2-u_1^2-u_2^2)\\
\\
I_2&=&-\dfrac{u_2^2(u_1^2-\kappa^2)p_{u_1}^2}{u_1^2-u_2^2}
-\dfrac{u_1^2(u_2^2-\kappa^2)p_{u_2}^2}{u_2^2-u_1^2}+\alpha^2u_1^2u_2^2\,.
\end{array}
\eq
Rewriting equation of motion
 \[
 \dfrac{d{u}_1}{dt}=\{u_1,H\}=\dfrac{(u_1^2-\kappa^2)p_{u_1}}{u_1^2-u_2^2}\,,\qquad
  \dfrac{d{u}_2}{dt}=\{u_2,H\}=\dfrac{(u_2^2-\kappa^2)p_{u_2}}{u_2^2-u_1^2}\,,
 \]
 in the following form
 \[
 \dfrac{du_1}{p_{u_1}}=\dfrac{dt}{u_1^2-u_2^2}\,,\qquad  \dfrac{du_2}{p_{u_2}}=-\dfrac{dt}{u_1^2-u_2^2}\,,
 \]
 we can eliminate time  and obtain equation
 \[
 \dfrac{du_1}{p_{u_1}} + \dfrac{du_2}{p_{u_2}}=0\,.
 \]
Substituting solutions of equations (\ref{int-osc1}) with respect to $p_{u_{1,2}}$ into this expression  and integrating  we obtain 
standard equation defining form of the trajectories  \cite{lag}:
\[
\int\dfrac{du_1}{\sqrt{(u_1^2-\kappa^2) (\alpha^2u_1^4+(I_1-\alpha^2 \kappa^2) u_1^2+I_2)}}+
\int\dfrac{du_2}{\sqrt{(u_2^2-\kappa^2) (\alpha^2u_2^4+(I_1-\alpha^2 \kappa^2) u_2^2+I_2)}}=const
\]
This equation is reduced to (\ref{ab-eq2}) using Euler's substitution $u_i^2=x_i$, which  allows us to consider  evolution of divisors on elliptic curve $X$
\bq\label{ell-osc}
X:\quad \Phi(x,y)=y^2-f(x)=0\,,\quad f(x)=\alpha^2x^4+(I_1-2\alpha^2\kappa^2)x^3+(\alpha^2\kappa^4-I_1\kappa^2+I_2)x^2-\kappa^2 I_2x
\eq
governed by equation (\ref{div-eq})
\[
P_1(t)+P_2(t)=P_3\,.
\]
Coordinates of movable points $P_{1,2}(t)$  are
\[x_1=u_1^2,\quad y_1=(u_1^2-\kappa^2)u_1p_{u_1}\qquad\mbox{and}\qquad x_2=u_2^2\,,\quad y_2=(u_2^2-\kappa^2)u_2p_{u_2}\,.\]
The abscissa of fixed point $x_3$ (\ref{coord-x3}) reads as
\[\begin{array}{c}
x_3=
-\frac{\bigl((u_1^2-\kappa^2)u_2p_{u_1}-(u_2^2-\kappa^2)u_1p_{u_2}+\alpha u_1u_2(u_1^2-u_2^2)\bigr)^2}{(u_1^2-u_2^2)
\bigl(
(\kappa^2-u_1^2)p_{u_1}^2-(\kappa^2-u_2^2)p_{u_2}^2+
2\alpha(u_1(\kappa^2-u_1^2)p_{u_1}-u_2(\kappa^2-u_2^2)p_{u_2})+
\alpha^2(\kappa^2-u_1^2-u_2^2)(u_1^2-u_2^2)
\bigr)
}\,,
\end{array}
\]
Ordinate $y_3$ (\ref{coord-y3}) is a more lengthy rational function in elliptic coordinates, which we do not present for brevity,  whereas Euler's integral (\ref{eul-int}) is the following simple  polynomial
\[
C=-(p_1q_2+p_2q_1)^2-\kappa^2 I_1+\alpha^2\kappa^4\,.
\]
Existence of the polynomial first integrals $I_1, I_2$ and $C$ is related to symmetries of equations of motion  in the original physical space. Existence of non-polynomial first integrals $x_3$ and $y_3$ is related to motion along elliptic curve $X$ around fixed point $P_3$.

Symmetry breaking transformation  (\ref{trans-mom}) generates polynomial integrals of motion
\bq\label{int-osc-mn}
 \begin{array}{rcl}
 I_1&=&\dfrac{u_1^2-\kappa^2}{u_1^2-u_2^2}\left(\dfrac {p_{u_1}}{m}\right)^2+
 \dfrac{u_2^2-\kappa^2}{u_2^2-u_1^2}\left(\dfrac {p_{u_2}}{n}\right)^2
  +\alpha^2(\kappa^2-u_1^2-u_2^2)\\
 \\
I_2&=&\dfrac{u_2^2(u_1^2-\kappa^2)}{u_2^2-u_1^2}\left(\dfrac {p_{u_1}}{m}\right)^2
 +\dfrac{u_1^2(u_2^2-\kappa^2)}{u_1^2-u_2^2}\left(\dfrac {p_{u_2}}{n}\right)^2+\alpha^2u_1^2u_2^2\,,
 \end{array}
 \eq
  In Cartesian coordinates on the  plane superintegrable   Hamiltonians in (\ref{int-osc-mn})  read as
\[
H=\dfrac{(m^2+n^2)(p_1^2+p_2^2)}{4 m^2 n^2}
+\dfrac{(m^2-n^2)\Bigl( (\kappa^2-q_1^2+q_2^2) (p_1^2-p_2^2)+4 q_1 q_2 p_1 p_2\Bigr)}{4 m^2 n^2 r r'}-\dfrac{\alpha^2(q_1^2+q_2^2))}{2}\,.
\]
where $r$, $r'$ and $\kappa$  enter into the definition of elliptic coordinates. The corresponding  first integrals $x_3, y_3$  (\ref{xy3-mn}) and $C_{mn}$ (\ref{eul-int-nm}) are rational functions at $m\neq n$. Some particular expressions for these first integrals may be found in \cite{ts19s}.

\begin{prop}
Functions  $I_1,I_2$ (\ref{int-osc-mn}) and $x_3,y_3$ (\ref{xy3-mn}) on $T^*\mathbb R^2$
can be considered as representation of the following algebra of the first integrals
\bq\label{alg-int-osc}
\begin{array}{lll}
 \{I_1,I_2\}=0\,,\qquad &\{I_1,x_3\}=0\,,\qquad &\{I_1,y_3\}=0\,,\\ \\
 \{I_2,x_3\}=2\Phi_y(x_3,y_3)\,,\qquad &\{I_2,y_3\}= -2\Phi_x(x_3,y_3)\,,\qquad &\{x_3,y_3\}=2x_3(\kappa^2-x_3^2)\,,
 \end{array}
\eq
labelled by two integer numbers $m$ and $n$.  Here
\[
\Phi_y(x,y)=2y\,,\qquad  -\Phi_x(x,y)
 =4\alpha^2x^3+3(I_1-2\alpha^2\kappa^2)x^2+2(\alpha^2\kappa^4-\kappa^2I_1+I_2) x-\kappa^2I_2
\]
are derivatives of  function $\Phi(x,y)$ from the definition of elliptic curve $X$ (\ref{ell-osc}), and $\{.,.\}$
is the standard canonical Poisson bracket.
\end{prop}
As in Section 2 algebra of the first integrals  (\ref{alg-int-osc})  is derived from the Poisson brackets between the corresponding action-angle variables. We also have computer-assisted proof of this Proposition at $m=1,2,3$ and $n=1,2,3$.

This algebra of the first integrals (\ref{alg-int-osc}) slightly differs from (\ref{alg-int-kepl}) because in the Kepler problem we take
$x_{1,2}=u_{1,2}$, whereas for oscillator we have to put $x_{1,2}=u_{1,2}^2$ and, therefore,  we have different Poisson brackets between coordinates of movable points.

\subsection{Smorodinsky-Winternitz system}
In order to obtain  the so-called Smorodinsky-Winternitz system \cite{sw}
we have to start with elliptic curve $X$ defined by equation $\Phi(x,y)=y^2-f(x)=0$ with
\bq\label{ell-sw}
 f(x)=
\alpha^2x^4+(I_1-2\alpha^2\kappa^2)x^3+
(\alpha^2\kappa^4-I_1\kappa^2+I_2-2\beta-\gamma)x^2-\kappa^2(I_2-4\beta+\delta)x
-2\beta\kappa^4
\eq
instead of (\ref{ell-osc}). Equation (\ref{div-eq-mn})
\[
[m]P_1(t)+[n]P_2(t)=P_3
\]
determines evolution of two moving points around a third fixed point in the intersection divisor.  Coefficients $I_{1,2}$ of polynomial $f(x)$ together with coordinates $(x_3,y_3)$ of fixed point $P_3$ are constants of the divisor motion.

Constants of the divisor motion give rise to the first integrals on phase space, which can be calculated using standard algorithm:
\begin{itemize}
  \item identify affine coordinates of movable points $P_ {1,2} (t) $ on the projective plane   with elliptic coordinates and momenta in phase space \[x_1=u_1^2,\quad y_1=(u_1^2-\kappa^2)u_1\dfrac{p_{u_1}}{m}\qquad\mbox{and}\qquad x_2=u_2^2\,,\quad y_2=(u_2^2-\kappa^2)u_2\dfrac{p_{u_2}}{n} \,; \]
  \item solve a pair of equations $\Phi(x_1,y_1)=0$ and $\Phi(x_2,y_2)$ with respect to  $I_1, I_2$; 
 \item calculate first integrals associated with affine coordinates $(x_3,y_3) $  (\ref{xy3-mn}) of the fixed point $P_3$\,.
\end{itemize}
After that we can verify that functions  $I_1,I_2$ and $x_3,y_3$ on $T^*\mathbb R^2$ satisfy to the Poisson brackets (\ref{alg-int-osc}).

For the curve $X$ (\ref{ell-sw} ) one gets the following Hamiltonian 
  \[H=\dfrac{I_1}{2}=T_{mn}- \dfrac{\alpha^2}{2}(q_1^2+q_2^2)+\dfrac{\beta}{q_1^2}+\dfrac{\gamma}{q_2^2}.\]
Here potential part is independent on integer numbers $m$ and $n$, whereas kinetic energy $T_{mn}$ is equal to
\[\begin{array}{rcl}
T_{mn}&=&\dfrac{u_1^2-\kappa^2}{u_1^2-u_2^2}\left(\dfrac {p_{u_1}}{m}\right)^2+
 \dfrac{u_2^2-\kappa^2}{u_2^2-u_1^2}\left(\dfrac {p_{u_2}}{n}\right)^2\\
 \\
 &=&\dfrac{(m^2+n^2)(p_1^2+p_2^2)}{4 m^2 n^2}
+\dfrac{(m^2-n^2)\Bigl( (\kappa^2-q_1^2+q_2^2) (p_1^2-p_2^2)+4 q_1 q_2 p_1 p_2\Bigr)}{4 m^2 n^2 r r'}\,,
\end{array}
\]
where $r$, $r'$ and $\kappa$  enter into  Euler's definition of elliptic coordinates on the plane. At $m=n=1$ this Hamiltonian coincides with the Hamiltonian of the Smorodinsky-Winternitz system \cite{sw}.

\section{Drach  system}
In 1935 Jules Drach classified  Hamiltonian systems in $T^*\mathbb R^2$  with third order integrals of motion  \cite{dr35}. Below we consider the so-called (h) Drach  system  associated with elliptic curve, see details of classification  in \cite{ts00,ts08,ts08a}. Possible generalizations of the Drach systems are discussed in \cite{ts11}.

The (h) Drach system is defined by Hamiltonian
\[
H=I_1=p_1p_2-2\alpha(q_1+q_2)-\beta\left(\dfrac{q_1}{2\sqrt{q_2}}+\dfrac{3\sqrt{q_2}}{2}\right)-\dfrac{\gamma}{2\sqrt{q_2}}
\]
 and first integral
 \[
 I_2=(q_1+q_2)p_1p_2-q_1p_1^2-q_2p_2^2-\alpha(q_1-q_2)^2-\dfrac{\beta(q_1-q_2)^2}{2\sqrt{q_2}}
 -\dfrac{\gamma(q_1-q_2)}{2\sqrt{q_2}}\,.
 \]
 After canonical point transformation of variables
 \[
 q_1 =\dfrac{(u_1-u_2)^2}{4}\,,\quad p_1 = \dfrac{p_{u_1}-p_{u_2}}{u_1-u_2}\,,\quad
 q_2 =\dfrac{(u_1+u_2)^2}{4}\,,\quad p_2 = \dfrac{p_{u_1}+p_{u_2}}{u_1+u_2}
 \]
 integrals of motion look like
 \[\begin{array}{rcl}
 I_1&=&\dfrac{p_{u_1}^2}{u_1^2-u_2^2}
 +\dfrac{p_{u_2}^2}{u_2^2-u_1^2}-\alpha(u_1^2+u_2^2)-\dfrac{\beta(u_1^2+u_1u_2+u_2^2)}{u_1+u_2}-\dfrac{\gamma}{u_1+u_2}
 \\ \\
  I_2&=&
 \dfrac{u_2^2p_{u_1}^2}{u_1^2-u_2^2}
 \dfrac{u_1^2p_{u_2}^2}{u_2^2-u_1^2}
 -\alpha u_1^2u_2^2-\dfrac{\beta u_1^2u_2^2}{u_1+u_2}
 +\dfrac{\gamma u_1u_2}{u_1+u_2}\,.
 \end{array}
 \]
Solving these equations with respect to $p_{u_1}$ and $p_{u_2}$ we obtain separated relations
\bq\label{ell-dr}
\Phi_i(u_i,p_{u_i})= p_{u_i}^2-\Bigl(\alpha u_i^4+\beta u_i^3+I_1 u_i^2+\gamma u_i-I_2\Bigr) \,,\qquad i=1,2.
\eq
Following  \cite{st95}  we determine St\"{a}ckel matrix $S$ with entries
\bq\label{st-mat}
S_{ij}=\dfrac{\partial \Phi_j}{\partial I_i}
\eq
and St\"{a}ckel angle variables
\[
\omega_1=\dfrac{1}{2}\int \dfrac{S_{11}du_1}{p_{u_1}}+ \dfrac{1}{2}\int \dfrac{S_{12}du_2}{p_{u_2}}\,,\quad
\omega_2=\dfrac{1}{2}\int \dfrac{S_{21}du_1}{p_{u_1}}+ \dfrac{2}{2}\int \dfrac{S_{22}du_2}{p_{u_2}}\,,\\
\]
which can be rewritten in a standard form for the St\"{a}ckel systems with $n$ degrees of freedom
\bq\label{angle-st-gen}
\omega_j=- \sum_{i=1}^n \int^{P_i} \dfrac{\partial \Phi(x,y)/\partial I_j}{\partial \Phi(x,y)/\partial y}\,dx\,,
\eq
using separated relations (\ref{ell-dr}), definition of the St\"{a}ckel matrix (\ref{st-mat} ) and definition of   points $P_i=(x_i,y_i)$ on hyperelliptic curve $X$, see \cite{ts99}.

In action-angle variables $I_{1,2}$ and $\omega_{1,2}$ equations of motion and symplectic form  look like
\[
\dot{I}_i=0\,,\qquad \dot{\omega}_i=\dfrac{\partial H}{\partial I_i}\,,\qquad \Omega=dI_1\wedge d \omega_1+dI_2\wedge d \omega_2\,.
\]
Because $H=I_1$, differential equations are trivially reduced to quadratures, for instance
\[
I_{1,2}=const,\qquad
\omega_2=-\dfrac{1}{2}\sum_{i=1}^2 \int^{P_i} \dfrac{dx_1}{\sqrt{\alpha x^4+\beta x^3+I_1x^2+\gamma x-I_2}}=const\,.
\]
Relation $\omega_2=const$  involves the sum of Abelian integrals with holomorphic differentials  on $X$ and, therefore, it defines swing of two points around a third fixed point on elliptic curve (\ref{div-eq})
\[ P_1(t)+P_2(t)=P_3\,.\]
In the Drach case coordinates of moving points are simple  function on physical variables
\[
x_1=u_1\,,\quad y_1=p_{u_1}\,,\qquad x_2=u_2\,,\quad y_2=p_{u_2}\,,
\]
and, therefore, abscissa of fixed point $P_3$ is a quite observable rational function
\[\begin{array}{l}
x_3=-\frac{
2(u_1^2-u_2^2)(u_2p_{u_1}-u_1p_{u_2})\sqrt{\alpha}+u_1u_2(u_1-u_2)^2\beta-(u_1-u_2)^2\gamma-2(p_{u_1}-p_{u_2})(u_2p_{u_1}-u_1p_{u_2})
}{\bigl(2(u_1^2-u_2^2)(u_1+u_2)\alpha-2(p_{u_1}-p_{u_2})\sqrt{\alpha}+(u_1-u_2)\beta\bigr)(u_1^2-u_2^2)}\,,
\end{array}
\]
similar to Euler's integral (\ref{eul-int}), which  is the following polynomial in momenta
\[
C=\dfrac{(p_{u_1}-p_{u_2})^2}{(u_1-u_2)^2}-(u_1+u_2)^2\alpha-(u_1+u_2)\beta\,.
\]

Let us apply symmetry breaking transformation  (\ref{trans-mom}) to this superintegrable St\"{a}ckel system.
Action variables $I_{1,2}$ associated with the equation
\[[m]P_1(t)+[n]P_2(t)=P_3\,,\]
are equal to
\bq\label{int-dr-mn}
\begin{array}{rcl}
 I_1&=&\dfrac{p_{u_1}^2/m^2}{u_1^2-u_2^2}
 +\dfrac{p_{u_2}^2/n^2}{u_2^2-u_1^2}-\alpha(u_1^2+u_2^2)-\dfrac{\beta(u_1^2+u_1u_2+u_2^2)}{u_1+u_2}-\dfrac{\gamma}{u_1+u_2}
 \\ \\
  I_2&=&
 \dfrac{u_2^2p_{u_1}^2/m^2}{u_1^2-u_2^2}+
 \dfrac{u_1^2p_{u_2}^2/n^2}{u_2^2-u_1^2}
 -\alpha u_1^2u_2^2-\dfrac{\beta u_1^2u_2^2}{u_1+u_2}
 +\dfrac{\gamma u_1u_2}{u_1+u_2}\,.
 \end{array}
 \eq
 In original Cartesian coordinates Hamiltonians in (\ref{int-dr-mn}) have the form
 \[\begin{array}{rcl}
 I_1&=&\dfrac{(m^2+n^2)p_1p_2}{2m^2n^2}+\dfrac{(m^2-n^2)(\sqrt{q_1}+\sqrt{q_2})^2(q_1p_1^2+q_2p_2^2)}
 {4n^2m^2(\sqrt{q_1q_2}+q_1)(\sqrt{q_1q_2}+q_2)}\\ \\ &-&2\alpha(q_1+q_2)-\beta\left(\dfrac{q_1}{2\sqrt{q_2}}+\dfrac{3\sqrt{q_2}}{2}\right)-\dfrac{\gamma}{2\sqrt{q_2}}\,.
 \end{array}
 \]
Following \cite{ran16}  we can say that these Hamiltonians describe motion of the body  with a position dependent mass. 
 
 The corresponding St\"{a}ckel angle variables
\[
\omega_1=\dfrac{m}{2}\int \dfrac{S_{11}du_1}{p_{u_1}}+ \dfrac{n}{2}\int \dfrac{S_{12}du_2}{p_{u_2}}\,,\quad
\omega_2=\dfrac{m}{2}\int \dfrac{S_{21}du_1}{p_{u_1}}+ \dfrac{n}{2}\int \dfrac{S_{12}du_2}{p_{u_2}}\,,\\
\]
involve a holomorphic differential on elliptic curve, which allows us to calculate coordinates of the fixed point using arithmetic equation (\ref{xy3-mn}).

At   $m=2$ and $n=1$  abscissa of fixed point $P_3$ remains a quite observable rational function if $\beta=\gamma=0$
\[
x_3=-u_2-(u_1^2-u_2^2)p_{u_1}\left(\frac{1}{2\sqrt{\alpha}u_1(u_1^2-u_2^2)+u_2p_{u_1}-2u_1p_{u_2}}
+\frac{1}{2\sqrt{\alpha}u_1(u_1^2-u_2^2)-u_2p_{u_1}+2u_1p_{u_2}}\right)\,.
\]
This expression was obtained using doubling  of point $P_1=(u_1,p_{u_1}/2)$ and addition (\ref{xy3-mn}) of points $[2]P_1$ and $P_2=(u_2,p_{u_2})$ on elliptic curve $X$.

At $m=3$ and $n=1$ abscissa  of fixed point $P_3$ is a bulky function even if $\beta=\gamma=0$
\[\begin{array}{rcl}
x_3&=&
-\dfrac{(u_2p_{u_1}+3u_1p_{u_2})(u_2p_{u_1}-3u_1p_{u_2})^2}{
3\sqrt{\alpha}\bigl((4u_1^2-u_2^2) p_{u_1}^2-6 u_1 u_2 p_{u_1} p_{u_2}-9 u_1^2p_{u_2}^2 \bigr)(u_1^2-u_2^2)}+\\ \\
&+&6\sqrt{\alpha}(u_1^2-u_2^2)u_1^2 p_{u_1}^2
\left(\dfrac{(u_1-u_2)^2}{A_-} - \dfrac{(u_1+u_2)^2 }{A_+}\right)\,,
\end{array}
\]
where
\[
A_+=\Bigl(
9\alpha u_1^2 (u_1^2-u_2^2)^2 +(u_2p_{u_1}-3u_1 p_{u_2}) \bigl((2 u_1+u_2)p_{u_1}+3u_1 p_{u_2}\bigr)
\Bigr) \bigl((2u_1+u_2)p_{u_1}+3u_1 p_{u_2}\bigr)
\]
and
\[
A_-=\Bigl(
9\alpha u_1^2(u_1^2-u_2^2)^2-(u_2p_{u_1}-3u_1p_{u_2})\bigl((2u_1-u_2)p_{u_1}-3u_1p_{u_2}\bigr)
\Bigr)\bigl((2u_1-u_2)p_{u_1} -3 u_1p_{u_2}\bigr)
\]
This expression was obtained using tripling of point $P_1=(u_1,p_{u_1}/3)$ and addition (\ref{xy3-mn}) of points $[3]P_1$ and $P_2=(u_2,p_{u_2})$ on elliptic curve $X$.

 \begin{prop}
Functions  $I_1,I_2$ (\ref{int-dr-mn}) and $x_3,y_3$ (\ref{xy3-mn}) on $T^*\mathbb R^2$
can be considered as representation of the following algebra of the first integrals
\bq\label{alg-int-dr}
\begin{array}{lll}
 \{I_1,I_2\}=0\,,\qquad &\{I_1,x_3\}=0\,,\qquad &\{I_1,y_3\}=0\,,\\ \\
 \{I_2,x_3\}=\Phi_y(x_3,y_3)\,,\qquad &\{I_2,y_3\}= -\Phi_x(x_3,y_3)\,,\qquad &\{x_3,y_3\}=1
 \end{array}
\eq
labelled by two integer numbers $m$ and $n$.  Here
\[
\Phi_y(x,y)=2y\,,\qquad  -\Phi_x(x,y)=4\alpha x^3+3\beta x^2+2I_1 x+\gamma
 \]
are derivatives of  function $\Phi(x,y)$ from the definition of elliptic curve $X$ (\ref{ell-dr}), and $\{.,.\}$
is the canonical Poisson bracket.
\end{prop}
This algebra is derived from the Poisson bracket between the corresponding action-angle variables.
We also  have  computer-assisted proof of this Proposition at $m=1,2,3$ and $n=1,2,3$.

So, the so-called (h) Drach system  belongs to a family of  two-dimensional superintegrable systems associated with
elliptic curves of the form $X:\Phi(x,y)=y^2-f^{(k)}(x)$, where
 \[\begin{array}{rclrcl}
  f^{(1)}(x)&=&\alpha x^4+\beta x^3+\gamma x^2+I_1x+I_2\,,\qquad f^{(2)}(x)&=&\alpha x^4+\beta x^3+I_2 x^2+\gamma x+I_2\,,\\
  \\
  f^{(3)}(x)&=&\alpha x^4+I_2x^3+\beta x^2+\gamma x+I_2\,,\qquad f^{(4)}(x)&=&I_1x^4+\alpha x^3+\beta x^2+\gamma x+I_2\,,\\
\end{array}
\]
and equation of motion
\[
[m]P_1(t)+[n]P_2(t)=P_3\,, \qquad m,n\in \mathbb Z\,.
\]
 For all these superintegrable systems algebra of the first integrals has the standard form  (\ref{alg-int-dr}) which directly follows from the Poisson brackets between action-angle variables.

 \section{3D superintegrable  St\"{a}ckel system on elliptic curve}
Let us consider St\"{a}ckel system associated with symmetric product $X\times X \times X$ of elliptic curve $X$ defined by  equation of the form
\bq\label{ell-3d}
X:\quad \Phi(x,y)=y^2-f(x)=0\,,\qquad f(x)=\alpha x^4+\beta x^3+I_1x^2+I_2x+I_3\,.
\eq
If we  identify  coordinates of the points on each copy of $X$ in $X\times X \times X$ with canonical coordinates in $T^*\mathbb R^3$
\[
x_1=u_1\,,\quad y_1=p_{u_1}\,,\quad x_2=u_2\,,\quad y_2=p_{u_2}\,,\quad x_3=u_3\,,\quad y_3=p_{u_3}\,,
\]
we obtain action variables $I_1,I_2$ and $I_3$ :
\bq
\label{int-3d}
\begin{array}{rcl}
I_1&=&\scriptstyle \frac{p_{u_1}^2}{(u_1-u_3)(u_1-u_2)}+\frac{p_{u_2}^2}{(u_2-u_3)(u_2-u_1)}
+\frac{p_{u_3}^2}{(u_3-u_1)(u_3-u_2)}- (u_1^2+u_2^2+u_3^2+u_1u_2+u_1u_3+u_2u_3)\alpha-(u_1+u_2+u_3)\beta\\ \\
I_2&=&
\scriptstyle -\frac{(u_2+u_3)p_{u_1}^2}{(u_1-u_3)(u_1-u_2)}-\frac{(u_1+u_3)p_{u_2}^2}{(u_2-u_3)(u_2-u_1)}
-\frac{(u_1+u_2)p_{u_3}^2}{(u_3-u_1)(u_3-u_2)}
+(u_1+u_2)(u_1+u_3)(u_2+u_3)\alpha+
(u_1u_2+u_1u_3+u_2u_3)\beta\\ \\
I_3&=&\scriptstyle
\frac{u_2u_3p_{u_1}^2}{(u_1-u_3)(u_1-u_2)}+\frac{u_1u_3p_{u_2}^2}{(u_2-u_3)(u_2-u_1)}
+\frac{u_1u_2p_{u_3}^2}{(u_3-u_1)(u_3-u_2)}
-u_1u_2u_3(u_1+u_2+u_3)\alpha -u_1u_2u_3\beta\,.
\end{array}
\eq
solving separated relations $\Phi(u_i,p_{u_i}, I_1,I_2,I_3)=0$ with respect to $I_1,I_2$ and $I_3$. Substituting solutions of  the same separated relations with respect to $p_{u_1},p_{u_2}$ and $p_{u_3}$ into the St\"{a}ckel definition  (\ref{angle-st-gen}) we get standard angle variables
\[\begin{array}{rcl}
\omega_1&=&\displaystyle -\int \dfrac{u_1^2du_1}{\sqrt{f(u_1)}}-\int \dfrac{u_2^2du_2}{\sqrt{f(u_2)}}-\int \dfrac{u_3^2du_3}{\sqrt{f(u_3)}}\,,\\ \\
\omega_2&=&\displaystyle-\int \dfrac{u_1du_1}{\sqrt{f(u_1)}}-\int \dfrac{u_2du_2}{\sqrt{f(u_2)}}-\int \dfrac{u_3du_3}{\sqrt{f(u_3)}}\,,\\ \\
\omega_3&=&\displaystyle-\int \dfrac{du_1}{\sqrt{f(u_1)}}-\int \dfrac{du_2}{\sqrt{f(u_2)}}-\int \dfrac{du_3}{\sqrt{f(u_3)}}\,.
\end{array}
\]
Equation of motion $\omega_3=const$ involves a holomorphic differential on elliptic curve and, therefore, it  is equivalent to arithmetic equation for divisors on $X$
\bq\label{div-eq-3}
P_1(t)+P_2(t)+P_3(t)=P_4\,.
\eq
This equation describes the swing of parabola $Y$
\[
Y:\quad y=\mathcal P(x)\,,\qquad \mathcal P(x)=b_2(t)x^2+b_1(t)x+b_0(t)
\]
around some  fixed point $P_4$ on $X$. Because four  points $P_1,P_2,P_3$ and $-P_4$ form an intersection divisor of $X$ and $Y$ we can calculate  three coefficients $b_2,b_1$ and $b_0$ by solving three equations
\[
 y_1=b_2x_1^2+b_1x_1+b_0\,,\quad y_2=b_2x_2^2+b_1x_2+b_0\,,\quad y_3=b_2x_3^2+b_1x_3+b_0\,.
\]
Substituting  $ y=\mathcal P(x)$ into  definition $y^2-f(x)=0$  of $X$ we obtain Abel's polynomial
\[
\psi(x)=f(x)-\mathcal P^2(x)=(a_4-b_2^2)(x-x_1)(x-x_2)(x-x_3)(x-x_4)\,.
\]
Evaluating coefficients of this polynomial we find coordinates of the fixed point
\[
x_4= -x_1-x_2-x_3-\dfrac{a_3-2b_1b_2}{a_4-b_2^2}\,,\quad y_4=-\mathcal P(x_4)\,,
\]
which are constants of divisor motion (\ref{div-eq-3}) on elliptic curve $X$.

The corresponding rational functions on  phase space $T^*\mathbb R^3$
\bq\label{coord-p4}
x_4=-u_1-u_2-u_3-\dfrac{\beta-2b_1b_2}{\alpha-b_2^2}\,,\qquad  y_4=-(b_2x_4^2+b_1x_4+b_0)\,,
\eq
where
\[\begin{array}{rcl}
b_2&=&\dfrac{(u_2-u_3)p_{u_1}+(u_3-u_1)p_{u_2}+(u_1-u_2)p_{u_3}}{(u_1-u_2)(u_1-u_2)(u_2-u_3)}\,,\\ \\
b_1&=&-\dfrac{(u_2^2-u_3^2)p_{u_1}-(u_3^2-u_1^2)p_{u_2}-(u_1^2-u_2^2)p_{u_3}}{(u_1-u_2)(u_1-u_2)(u_2-u_3)}\,,\\ \\
b_0&=&\dfrac{u_2u_3(u_2-u_3)p_{u_1}+u_1u_3(u_3-u_1)p_{u_2}+u_1u_2(u_1-u_2)p_{u_3}}{(u_1-u_2)(u_1-u_2)(u_2-u_3)}\,,
\end{array}
\]
are first integrals of the dynamical system determined by Hamiltonian $H(I_1,I_2,I_3)$ and canonical Poisson brackets.

After symmetry breaking transformation  (\ref{trans-mom})
\[
p_{u_1}\to \dfrac{p_{u_1}}{m}\,,\qquad p_{u_2}\to \dfrac{p_{u_2}}{n}\,,\qquad p_{u_3}\to \dfrac{p_{u_3}}{k}
\]
equation of motion (\ref{div-eq-3}) on $X$ becomes
\[
[m]P_1(t)+[n]P_2(t)+[k]P_3(t)=P_4\,, \qquad m,n,k\in \mathbb Z\,.
\]
Affine coordinates of the constant part of intersection divisor are given by
\[
x_4= -[m]x_1-[n]x_2-[k]x_3-\dfrac{a_3-2b_1b_2}{a_4-b_2^2}\,,\quad y_4=-\mathcal P(x_4)\,,
\]
where parabola $Y: y=\mathcal P(x)$ is now defined by using Lagrange interpolation by movable points $[m]P_1$, $[n]P_2$ and $[k]P_3$ on elliptic curve $X$, where   \[P_1=(u_1,p_{u_1}/m)\,,\qquad P_2=(u_2,p_{u_2}/n)\,,\qquad P_3=(u_3,p_{u_3}/k)\,.\]

\begin{prop}
Functions  $I_1,I_2,I_3$ (\ref{int-3d}) and $x_4,y_4$ (\ref{coord-p4}) in phase space $T^*\mathbb R^3$
can be considered as representation of the following algebra of the first integrals
\bq\label{alg-int-3d}
\begin{array}{llll}
 \{I_1,I_2\}=0\,,\quad &\{I_1,I_3\}=0&\{I_1,x_3\}=0\,,\quad &\{I_1,y_3\}=0\,,\\ \\
 \{I_2,I_3\}=0\,,\quad &\{I_2,x_3\}=0\,,\quad &\{I_2,y_3\}=0\,,\\ \\
 \{I_3,x_4\}=\Phi_y(x_4,y_4)\,,\quad &\{I_3,y_4\}= -\Phi_x(x_4,y_4)\,,\quad &\{x_4,y_4\}=1
 \end{array}
\eq
Here
\[
\Phi_y(x,y)=\dfrac{\partial \Phi(x,y)}{\partial y}=2y\quad\mbox{and}\quad
\Phi_x(x,y)=\dfrac{\partial \Phi(x,y)}{\partial x}=-(4\alpha x^3+3\beta  x^2+2I_1 x+I_2)
  \]
are derivatives of  function $\Phi(x,y)$ from the definition of elliptic curve $X$ (\ref{ell-3d}) and $\{.,.\}$
is the canonical Poisson bracket.
\end{prop}
This algebra is derived from the Poisson bracket between the corresponding action-angle variables.
We also have  computer-assisted proof of this Proposition at $m=n=k$.

Algebra of the first integrals (\ref{alg-int-3d}) slightly differs from the algebra (\ref{alg-int-kepl}) in the Kepler case.  Abel's subalgebra of  (\ref{alg-int-3d})  consists of two elements $I_1$ and $I_2$, whereas Abel's subalgebra of  (\ref{alg-int-kepl}) has only one central element $I_1$.

 Summing up,  we can construct six families of  superintegrable systems
using  elliptic curves of the form $X:\Phi(x,y)=y^2-f^{(k)}(x)$, where
 \[\begin{array}{rclrcl}
  f^{(1)}(x)&=&\alpha x^4+\beta x^3+I_1x^2+I_2x+I_3\,,\qquad f^{(2)}(x)&=&\alpha x^4+I_1 x^3+\beta x^2+I_2x+I_3\,,\\
  \\
  f^{(3)}(x)&=&\alpha x^4+I_1x^3+I_2x^2+\beta x+I_3\,,\qquad f^{(4)}(x)&=&I_1x^4+\alpha x^3+\beta x^2+I_2x+I_3\,,\\
  \\
  f^{(5)}(x)&=&I_1x^4+\alpha x^3+I_2 x^2+\beta x+I_3\,,\qquad f^{(6)}(x)&=&I_1x^4+I_2x^3+\alpha x^2+\beta I_2x+I_3\,,
\end{array}
\]
and intersection divisor equation of motion
\[
[m]P_1(t)+[n]P_2(t)+[k]P_3(t)=P_4\,, \qquad m,n,k\in \mathbb Z\,.
\]
 For all these superintegrable systems algebra of the first integrals has the standard form  (\ref{alg-int-3d}) which directly follows from the Poisson brackets between action-angle variables. Following \cite{ran16}  we can say that Hamiltonians $I_1,I_2$ and $I_3$ describe motion of the body in $T^*\mathbb R^3$ with a position dependent mass.

\section{Conclusion}
Equations of motion
\[
\dot{z}_i=\{H,z_i\}\,,\qquad z_i\in T^*\mathbb R^n
\]
for the St\"{a}ckel systems on a symmetrized product  $X\times \cdots\times X$ of  hyperelliptic curve $X$
are equivalent to equation of motion
\[
\mbox{div} X\cdot Y(t)=0
\]
describing evolution of   the intersection divisor of $X$ and axillary curve $Y(t)$. For superintegrable St\"{a}ckel systems, intersection divisor  can be  divided on moving and fixed parts
 \[
\mbox{div} X\cdot Y(t)=D(t)+D'=0,
\]
 according to Abel's theorem. It is clear, that  constants of divisor motion are the coordinates of fixed part $D'$ of the intersection divisor, and   integrals of motion in phase space are some functions on these constants of divisor motion.  So, algebra of the first integrals in phase space can be obtained from the algebra of constants of divisor motion, which is easily obtained from the Poisson brackets between  the St\"{a}ckel action-angle variables.

In this note we calculate algebra of constants of  divisor motion  associated with the Kepler problem, harmonic oscillator, Drach system, St\"{a}ckel systems with two and three degrees of freedom and some of their deformations associated with symmetry breaking transformations of the St\"{a}ckel matrices. All these systems are related to various elliptic curves, but we can rewrite the corresponding algebras of non-polynomial integrals in a common form.

Scalar multiplication of points on elliptic curves
\[
\varphi:\quad X\to X\,,\qquad \varphi(P)= [m]P\,,
\]
 generates non-canonical transformation on phase space
 \[
 \psi:\quad  T^*\mathbb R^n\to T^*\mathbb R^n\,,\qquad  \psi(u_i)=u_i\,,\quad \psi(p_{u_i})= m_i p_{u_i}\,,
\]
which changes  the form of Hamiltonian, but  preserves its superintegrability property. Because multiplication of points on $X$ is a special case of isogenies between elliptic curves, we suppose that isogeny arithmetics also generates non-canonical transformations of phase space
\[\psi:\qquad T^*\mathbb R^n \to T^*\mathbb R^n\]
preserving superintegrability.  If this conjecture is true, than  isogeny volcanoes \cite{suz} could generate  superintegrable system  volcanoes.  In a forthcoming publication, we will discuss this conjecture and  application of   V\'{e}lu's formulas \cite{sil,velu} to construction superintegrable systems associated with  elliptic curves.

For superintegrable St\"{a}ckel systems on hyperelliptic curves of genus two we  have affine coordinates of divisors,  Mumford's coordinates of divisors, modified Jacobian coordinates,  Chudnovski-Jacobian coordinates, mixed coordinates, etc.  First integrals  associated with these coordinates could be algebraic, rational or polynomial functions in phase space satisfying various polynomial or non-polynomial relations.  It is interesting to study these relations associated with various constants of the divisor motion.

Class of superintegrable or degenerate systems is closely related to the class of bi-Hamiltonian systems with equations of motion
\[
\dfrac{d}{dt} z_i=\{I_1,z_i\}=\{I_2,z_i\}'\,,
\]
see \cite{ts15} and references within. So, we have two algebras of divisor motion constants  the  with respect to compatible Poisson brackets $\{.,.\}$ and $\{.,.\}'$. We suppose that both algebras of the first integrals are similar to "Hamiltonian equation of motion" with respect to "Hamiltonian" $\Phi(x,y)$ (\ref{eq-mn-ham}).

The work was supported by the Russian Science Foundation  (project  18-11-00032).

\end{document}